\documentclass[twocolumn,preprintnumbers,amsmath,amssymb,prb,aps]{revtex4}
\usepackage{txfonts}
\usepackage{pifont}
\usepackage{amssymb}
\usepackage{dcolumn}
\usepackage{amsmath}
\usepackage[dvips]{epsfig}

\makeatletter
\def\btt#1{\texttt{\@backslashchar#1}}%
\DeclareRobustCommand\bblash{\btt{\@backslashchar}}%
\makeatother

\begin{document}
\title{Orbital density wave order and electronic correlation driven insulating
$1$T-TaS$_{2}$ monolayer}
\author{Xiang-Long Yu$^{1}$}
\author{Da-Yong Liu$^{1}$}
\author{H.-Q. Lin$^{2}$}
\author{Ting Jia$^{1}$}
\author{Liang-Jian Zou$^{1,3}$}
\email[Corresponding author. E-mail: ]{zou@theory.issp.ac.cn}
\affiliation{1 Key Laboratory of Materials Physics,
              Institute of Solid State Physics, Chinese Academy of Sciences,
               P. O. Box 1129, Hefei 230031, China}
\affiliation{2 Beijing Computational Science Research Center, Beijing 100084, China}
\affiliation{3 Department of Physics,
                University of Science and Technology of China, Hefei 230026, China}
\date{\today}
\begin{abstract}
We present the orbital resolved electronic properties of structurally distorted
$1$T-TaS$_{2}$ monolayers. After optimizing the crystal structures, we obtain
the lattice parameters and atomic positions in the star-of-David structure, and
show the low-temperature band structures of distorted bulk are consistent with
recent angle resolved photoemission spectroscopy (ARPES) data. We further clearly
demonstrate that $5d$ electrons of Ta form ordered orbital-density-wave (ODW) state with
dominant $5d_{3{z}^2-{r}^2}$ character in central Ta, driving the one-dimensional
metallic state in paramagnetic bulk and half-filled insulator in monolayer.
Meanwhile, the star-of-David distortion in monolayers favors charge density wave
and the flat band stabilizes ferromagnetic density wave of Ta spins with the same
wavevector of ODW $\frac{4}{13}\mathbf{{b}_{1}}+\frac{1}{13}\mathbf{{b}_{2}}$.
We propose that $1$T-TaS$_{2}$ monolayer may pave a new way to
study the exciton physics, exciton-polaron coupling, and potential
applications for its exciton luminescence.

\end{abstract}

\maketitle


Layered transition-metal dichalcogenide $1$T-TaS$_{2}$ not only undergoes
several complicated charge-density-wave (CDW) phase transitions with decreasing
temperature, but also exhibits marvelous metal-insulator transition when
temperature further decreases to enough low T, and even becomes superconducting
upon doping \cite{AP24.117} or under high pressure \cite{PRB87.125135,NM7.960}.
To account for the nature of the low-T insulating phase in CDW ordered bulk
$1$T-TaS$_{2}$, the Mott insulator mechanism induced by electron-electron
correlation \cite{AP24.117,PMB39.229,PB99.183,RMP57.287} and the CDW insulator
mechanism arising from the electron-phonon coupling or Fermi surface nesting
\cite{PRB12.2220,PB99.188,PRB45.1462R} were proposed and have been debated
more than thirty years. Early argument that low-T $1$T-TaS2 is a Mott insulator
was based on the fact that a half-filled band crosses the Fermi energy and the
low-T resistivity diverges, thus the electronic correlation in $1$T-TaS$_{2}$
was believed to play crucial role in $1$T-TaS$_{2}$ in the low-T insulating
regime of T$<180$K \cite{PMB39.229,PB99.183}. However, the typical characters
of Mott insulator were not well confirmed experimentally until now:
lower and upper Hubbard bands and Mott insulating gap in optical spectra and
conductivity, magnetic moment and Curie-Weiss susceptibility in neutron scattering
and other magnetic experiments. Recent years the interest to nature of low-T phase
in $1$T-TaS$_{2}$ has revived \cite{NM7.2318,PRL109.176403,ARXIV1401.0246}
to examine the concept of Mott physics in distorted CDW phase, and tried to
find more clues in electronic states to confirm the role of electronic
correlation in the low-T insulating $1$T-TaS$_{2}$.

On the other hand, with the discovery of graphene, many experimentalists are
trying to exfoliate many layered/quasi-two-dimensional compounds, including
transition-metal dichalcogenides in order to find more monolayer systems,
and explore their unusual properties of these monolayer compounds/two-dimensional
electron systems \cite{PNAS102.10451}. One of recent successful examples is
the monolayer molybdenum dichalcogenides MoS2, in which Mo ions also form
a 2-dimensional hexagonal lattice, similar to graphene.
Despite of great works on two-dimensional MoS$_{2}$, the lack of high-quality
$1$T-TaS$_{2}$ sample hampers the study on monolayer $1$T-TaS$_{2}$.
Very recently, Chen {\it et al.} successfully exfoliated $1$T-TaS$_{2}$ thin films
of a few atomic layers experimentally, and observed a clear
transition from a metal to an insulator when the bulk reduces to several atomic
layers, and the CDW hysteresis gradually vanishes one by one when the thickness
of the thin film reduces to about 3 atomic layers\cite{PriCom}. Meanwhile, Darancet
{\it et al.} \cite{ARXIV1401.0246} showed that 1T-TaS$_{2}$ monolayer may be an
Mott insulator with a half-filled flat band.
These experimental and theoretical explorations may not only pave a new way to
study novel properties of 1T-TaS$_{2}$ monolayer with CDW state and Mott physics,
but also provide the possible applications of transition-metal dichalcogenides in
electronic devices and sensors.

%
%

In this Letter, we unambiguously demonstrate that the ground state of $1$T-TaS$_{2}$
monolayer in optimized distorted structure forms orbital order or orbital density
wave (ODW) states with the wavevector  $\frac{4}{13}\mathbf{{b}_{1}}+\frac{1}{13}\mathbf{{b}_{2}}$
 in monolayer $1$T-TaS$_{2}$.
In this scenario, the central Ta has dominant $5d_{3z^{2}-r^{2}}$ orbital component,
and orbital polarization gradually decays from the central Ta of star-of-David to
outer-ring Ta, we also predict the ferromagentic spin density wave state and distribution
of Ta magnetic moments, in consistent with the ODW in the distorted crystal
structures. We could also present the band structures of monolayer and few-layer
thin films, and point out that our calculated band structures agree with recent angle
resolved photoemission spectra (ARPES) data in distorted bulk. Such an ODW scenario
could interpret not only the quasi-one-dimensional metallic ground state in bulk
$1$T-TaS$_{2}$, but also predict the MIT when the system reduces to multilayer thin
film, and monolayer. We suggest that both the Coulomb correlation and electron-phonon
coupling play key roles.

We firstly perform the structural optimizations of $1$T-TaS$_{2}$ including monolayer,
and bulk both in the low-temperature phase with lattice
distortion and in the high-temperature phase without distortion. We find that
comparing with $1$T-TaS$_{2}$ bulk, the radius of star-of-David and the Ta-Ta
distance in monolayer slightly expand less than 0.3\%; on the other hand, the
Ta-S distance slightly shrinks, no more than 0.3\%.
From these optimized structures one obtains the basic features of the electronic
structure in high-symmetry normal phase and low-temperature distorted phase
through performing the generalized gradient approximation (GGA) calculations
and its correlation correction (GGA$+U$) for $1$T-TaS$_{2}$ bulk and
monolayer. The details of numerical calculations could be seen in the {\it Supplementary
Materials}.

The low-T distorted superstructure phase of bulk $1$T-TaS$_{2}$ is a commensurate
charge density wave (CCDW) state below $180$ K, and is characterized by
star-shaped clusters of $13$ atoms in the Ta plane \cite{AP24.117}.
The electronic band structures of bulk $1$T-TaS$_{2}$ without correlation correction
in the CCDW phase show that each star-of-David unit cell hosts one half-filled band crossing the Fermi
level with the energy bandwidth about $0.4$ eV, and this band displays a
rather weak in-plane dispersion but a strong out-of-plane one, indicating a
one-dimensional metal for bulk. Also there are a few bands from $-0.2$ eV to
$-1$ eV where only one or two bands are involved in high-T phase.
\begin{figure}[tp]
{\epsfig{figure=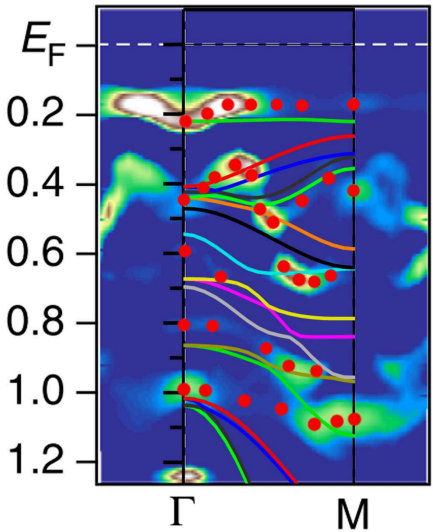,width=6.0cm,angle=0}}
\caption{(Color online) Comparison of bulk CCDW-state energy bands between our
results with the correlation correction of $U=2$ eV and ARPES experiment by
Ang {\it et al.}\cite{PRL109.176403}. Red points are artificially to
highlight occupied states in ARPES, and the curves are the energy bands
obtained by our first-principles calculations.}
\label{fig:bulk_band_ARPES}
\end{figure}
Moreover, with consideration of correlation correction, an interlayer-antiferromagnetic ground state
is obtained, and the half-filled band splits into upper and lower Hubbard bands.
These theoretical results are consistent with the available angle
resolved photoemission spectroscopy (ARPES) data, despite of dispersed results in the ARPES
experiments on the band structure and Fermi surface in the CCDW phase around
the $\Gamma$ point\cite{PRL81.1058,PRB64.245105,PRL107.177402,PRL109.176403}.
%
Comparing the details of our band structures with $U=2$ eV to recent ARPES results measured at
$30$ K by Ang $et\ al.$ along $\Gamma-M$\cite{PRL109.176403}, as shown
in Fig. \ref{fig:bulk_band_ARPES}, we find their global agreement in the energy
range from $E_{f}$ to $-1$ eV: near $-0.2$ eV, both experiment and calculation
results show flat dispersion and the maximum energy difference is about $0.05$ eV.
Below $-0.2$ eV, the momentum dependence of each band is in substantial agreement
with the ARPES data. In particular, considering rough bands degeneracy, our six bands around
$\Gamma$ point agree with Pillo {\it et al.}'s observation of the existence of at least six different
CDW-induced peaks \cite{PRB64.245105}.

Due to weak van der Waals interaction between TaS$_{2}$ layers, similar to graphene
and MoS$_{2}$, one may exfoliate a few layers $1$T-TaS$_{2}$, or even monolayer from
a bulk. For a 1T-TaS$_{2}$ monolayer, after optimizing its lattice structure and atomic
positions,
\begin{figure}[tp]
{\epsfig{figure=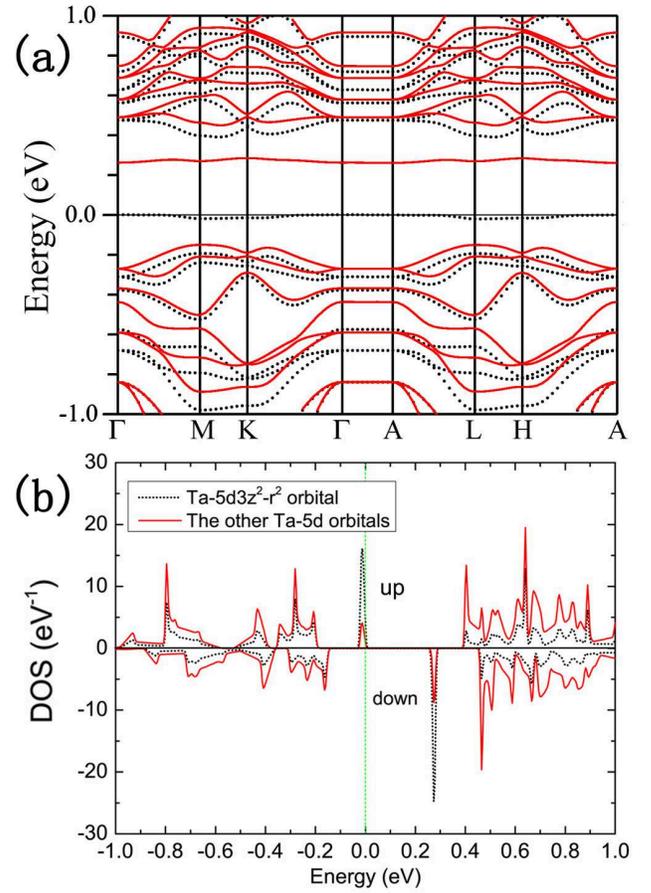,width=8.5cm,angle=0}}
\caption{(Color online) Band structures (a) and orbital resolved density of states (b) for distorted
monolayer $1$T-TaS$_{2}$ in the GGA $+ U$ scheme. Theoretical parameter $U=2$ eV.}
\label{fig:1slab_star_band_noso}
\end{figure}
our GGA calculation shows that a very narrow energy band with the bandwidth of less than $20$ meV
lies on the Fermi level E$_{F}$.
The evolution of such a flat band on E$_{F}$ with increasing distortion caould be seen in the
{\it Supplementary Materials}.
The broken of the translation invariance in the c-axis
in the monolayer leads to the vanish of the dispersion along $\Gamma - A$
direction, leaving a half-filled narrow band with significant contribution of
Ta$-5d_{3{z}^2-{r}^2}$ orbital. After taking into account the correlation correction
in the GGA$+U$ calculation with $U=2$eV, the half-filled flat band on E$_{F}$
splits into empty upper Hubbard band and
a full-filled lower Hubbard band, opening a gap of $\sim0.25$ eV,
as shown in Fig. \ref{fig:1slab_star_band_noso} (a).
Further the DOS in Fig. \ref{fig:1slab_star_band_noso} b displays that
the orbital symmetry character of DOS around E$_F$ are major Ta 5d$_{3z^{2}-r^{2}}$,
and the ground state of 1T-TaS$_{2}$ monolayer is orbital and spin polarized.

In $1$T-TaS$_{2}$, due to the trigonal crystalline field of S ions on Ta ions
in TaS$_{6}$ octahedra, the active orbitals of Ta 5d electrons are three
t$_{2g}$ components d$_{3{z}^2-{r}^2}$, e$_{g1}$ and e$_{g2}$, which are
the recombination of three orbital components d$_{XY}$, d$_{YZ}$ and d$_{XZ}$
(here X, Y, and Z are the variables in the original global coordinate system)
in accordance with the trigonal crystalline field symmetry.
The orbital character of the energy band around the Fermi level is also
analyzed quantitatively, and the orbital weights are summarized in Table I.
Ta $5d_{3{z}^2-{r}^2}$ orbital contributes significantly at $E_{F}$, especially
the one of Ta atoms in the interior of the star. The weights of the other
four 5d orbitals are nearly negligible. Such an orbital feature not only
address the unique dispersion in the $ab$-plane, but also is consistent with
the one-dimensional metallicity along c direction in bulk $1$T-TaS$_{2}$.
It is thus naturally expected that the metal-insulator transition occurs
when the bulk reduces to a monolayer.
It is worth noting that the weight of the $5d_{3{z}^2-{r}^2}$ orbital decreases
gradually from the center to the edge in the Ta star, forming an ODW. This unusual behavior
origins from the star-of-David distortion. Besides the inward contraction of
Ta atoms, the out-of-plane buckling of S atoms also plays an important role.

\begin{figure}[tp]
{\epsfig{figure=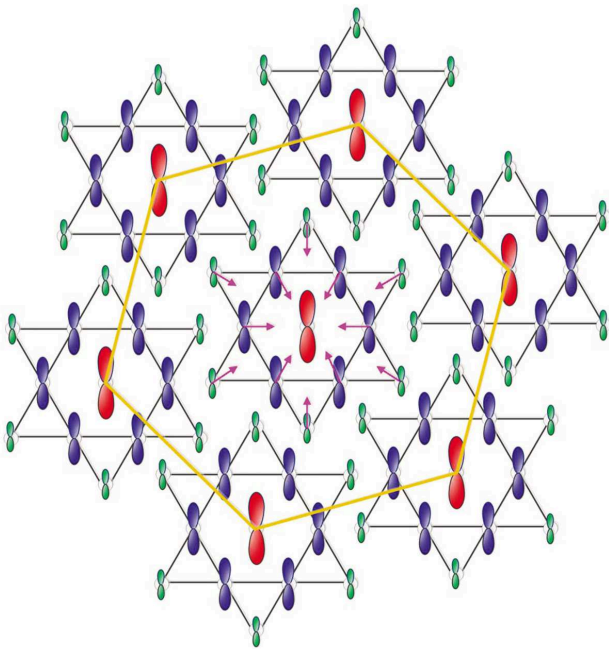,width=8.5cm,angle=0}}
\caption{(Color online) Schematic diagram of the orbital density wave order
in a star of Ta lattice in $1$T-TaS$_{2}$ monolayer. Arrows indicate the
distortion direction of Ta atoms. Ferromagnetic spin
density wave shares the same period.}
\label{fig:ODW}
\end{figure}

Fig. \ref{fig:ODW} is a schematic diagram of the atomic position and orbital
distribution in a star: the Ta-Ta distance of the internal ring is shorter than
that of outer ring so that Ta-S-Ta angle is small and height of Ta-S is large,
which is in favor of a hybridization between Ta$-5d$ and S$-4p_{z}$ orbital and
contributes major $5d_{3{z}^2-{r}^2}$ orbital component; as a contrast, in the
outer ring of stars long distance between two Ta atoms, large Ta-S-Ta angle
and short vertical height of Ta-S atoms lead to a weaken hybridization of
Ta$-5d$ and S$-4p_{z}$ orbitals, while an enhanced Ta$-5d$ and
S$-4p_{x/y}$ hybridization; thus from the center to the edge of star,
the 5d$_{3{z}^2-{r}^2}$ orbital polarization gradually reduces; therefore,
the orbital polarization is upmost in the center of star, and minimum in
the edge, forming a ferro-orbital density wave order over the whole
monolayer. The distribution of orbital polarization in a Ta star is summarized
in Table I.
One anticipates that in a bulk such an orbital order scenario
also validates.
%

It is interesting that whether CDW order remains exist in $1$T-TaS$_{2}$ monolayer.
From our optimized lattice one notices that the lattice distortion in monolayer
is still significant and has a little difference from the CCDW phase in bulk.
However, a very recent experiment by Chen {\it et al.} \cite{PriCom} showed that
the thermal hysteresis in resistivity indicating the pinning of CDW greatly weakens
and even disappears once $1$T-TaS$_{2}$ bulk reduces to a few layers.
The Ta plane shown in Fig. \ref{fig:ODW} illustrates the CDW-induced
displacements that lead to the commensurate $\sqrt{13}\times \sqrt{13}-R13.9{}^\circ $
superlattice, similar to observed in the low-temperature phase of bulk\cite{PM31.255,JPSJ53.1103,NM7.960}.
The corresponding lattice vectors of the superlattice with respect to that in high-temperature
homogeneous phase are $\mathbf{{{R}_{1}}}=3\mathbf{{{a}_{1}}}+\mathbf{{{a}_{2}}},\ \mathbf{{{R}_{2
}}}=-\mathbf{{{a}_{1}}}+4\mathbf{{{a}_{2}}}$.

\begin{figure}[tp]
{\epsfig{figure=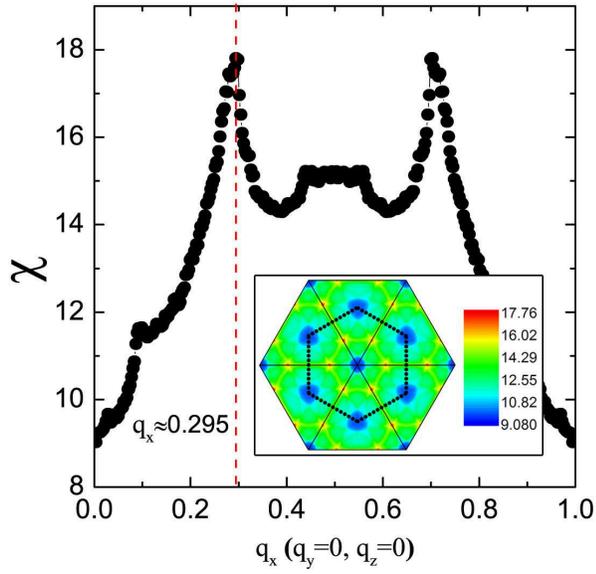,width=8.0cm,angle=0}}
\caption{(Color online) Real part of static electronic susceptibility with
arbitrary unit $vs$ wavevector along $\Gamma - M$ direction. Inset shows its
distribution in momentum space. The dotted line frame corresponds to the first
Brillouin zone of the monolayer in high-temperature normal phase.}
\label{fig:sus_noso_y0_z0}
\end{figure}

To indicate the role of electronically driven
instabilities as the origin of the star-of-David distortion in the ground state,
we have also analyzed the Fermi surface nesting in high-temperature homogeneous
phase and calculated the static electronic susceptibility
$\chi$, which is defined as
\begin{align}
\chi \left( \mathbf{q} \right)=\frac{1}{N}\sum\limits_{k,m,n}{\frac{f\left( {
{\varepsilon }_{n}}\left( \mathbf{k} \right) \right)\left[ 1-f\left( {{
\varepsilon }_{m}}\left( \mathbf{k}+\mathbf{q} \right) \right) \right]}{{{
\varepsilon }_{m}}\left( \mathbf{k}+\mathbf{q} \right)-{{\varepsilon }_{n}}
\left( \mathbf{k} \right)+i\eta }}.
\end{align}
Here ${\varepsilon }_{m}({\bf k})$ is the quasiparticle spectrum of the energy
band $m$ obtained from the first-principles calculations by the GGA method,
from which one may get insight to the charge and magnetic instability of
monolayer $1$T-TaS$_{2}$.
The electronic susceptibility in $1$T-TaS$_{2}$ monolayer shown in Fig.
\ref{fig:sus_noso_y0_z0} indicates a sharp peak, corresponding to a divergence,
implying that the Fermi surface nesting of the monolayer is perfect.
The global maximum of the electronic susceptibility appears around
$0.295\mathbf{b_{1}}$, corresponding to a nesting vector.
However, through further analysis, we find the calculated nesting vectors and
the star-of-David displacements do not match very well.
As shown above, the electronic susceptibility displays two similar but independent
nesting vectors
$\mathbf{{{Q}_{1}}}=0.295\mathbf{{{b}_{1}}}$ and
$\mathbf{{{Q}_{2}}}=0.295\mathbf{{{b}_{2}}}$,
respectively. A possible spin and charge density waves may oscillate as
$\cos \left( {{\mathbf{Q}}}\cdot \mathbf{R} \right)$ on the Ta plane, with a
wave vector $\mathbf{Q}$. This vector is equal to a commensurate
linear combination of $\mathbf{{{Q}_{1}}}$ and $\mathbf{{{Q}_{2}}}$, namely
$\mathbf{Q}=x\mathbf{{{Q}_{1}}}+y\mathbf{{{Q}_{2}}}$.
Through detailed calculations, a wave vector $\mathbf{Q}=\frac{4}{13}\mathbf{{{b}_{1}}}
+\frac{1}{13}\mathbf{{{b}_{2}}}$ is determined with $x=1.043$ and $y=0.261$,
consistent with the star-of-David distortion\cite{PM31.255,NM7.960,Science344.177}.

As a strong e-ph-coupling Mott insulator, the magnetic properties of $1$T-TaS$_{2}$
bulk remains an unsettled problem. Earlier data showed that CCDW bulk is
paramagnetic \cite{AP24.117,PMB39.229,PB99.183}, whereas, recent experiments
suggested a strong layered ferromagnetic order \cite{PRB71.153101}. Revealing the
magnetism of $1$T-TaS$_{2}$ monolayer not only favors but also is useful for
our understanding the magnetism of $1$T-TaS$_{2}$ bulk. Our correlation correction
calculation in the GGA$+U$ scheme shows
that magnetic moment is mainly contributed from the central and first-ring
Ta $5d_{3{z}^2-{r}^2}$ orbitals of the star cluster. The energy gap opening causes
a nonuniform distribution of magnetic moment in the star cluster where
the magnetic moments per Ta atom for the central, first-ring and
second-ring Ta are $0.272$, $0.04$ and $0.01 {{\mu }_{B}}$, respectively,
demonstrating that the ground state of a monolayer $1$T-TaS$_{2}$ is
ferromagnetic density wave ordered. The distribution of magnetic moments of Ta in
a star-of-David is also summarized in Table I.


\begin{figure}[tp]
{\epsfig{figure=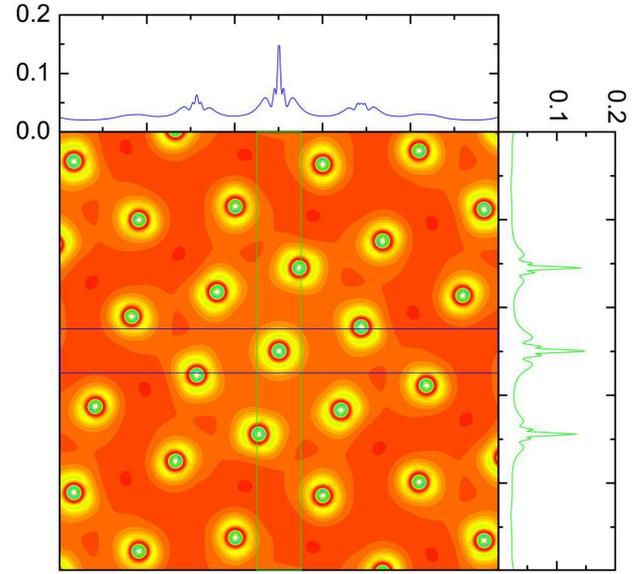,width=8.5cm,angle=0}}
\caption{(Color online) Charge density distribution of Ta-atom cross section
in distorted monolayer $1$T-TaS$_{2}$. The top blue and right green lines
correspond to the charge density in the blue and green boxes, respectively.}
\label{fig:star_char_den}
\end{figure}

For an intuitive picture of the CCDW phase, the charge density has been calculated
with cutoff energy -1Ry to truncate the contribution from atomic core.
Fig. \ref{fig:star_char_den} displays the charge density of Ta-atom cross section.
One can find that every 13 Ta atoms form a star-of-David cluster with a perfect
six-fold symmetry, and the charge density of the interval regime between two clusters
is small, just
like a moat which leads to an in-plane insulator. This is also consistent with our
preceding band analysis. These interesting and unusual behaviors of 1T-TaS$_{2}$
monolayer all originate from the formation of the star-of-David distortion and the ODW order.


We have shown that orbital physics plays a key role in the $5d$ $1$T-TaS$_{2}$
monolayer, as well as bulk. Though on-site Coulomb interaction may be
necessary and the Mott localization may happen in two-dimensional monolayer, such
as TiS$_{2}$ and VS$_{2}$, we attribute the origin of the orbital density wave
mainly to strong e-ph coupling in $1$T-TaS$_{2}$, since we find that sufficient
large Coulomb correlation $U$ could not drive the high-T undistorted phase of a
monolayer from metal to insulator, implying the e-ph coupling driven Jahn-Teller-like
distortion establish predominant $5d_{3{z}^2-{r}^2}$ orbital component and
the ferro-orbital density wave order from a undistorted three-orbital system.
Therefore when a $1$T-TaS$_{2}$ bulk reduces to a few
layers or even monolayer, conduction electrons in $d_{3z^{2}-r^{2}}$ orbital become
discrete and localized. The Coulomb interaction and e-ph coupling drive the
ferromagnetic and charge density waves in the monolayer with the same wavevector of
the orbital order.

From our study for various situations we find that the $LS$ coupling do not play considerable
roles on the ground state. This is a resultant of the orbital order in TaS$_{2}$ monolayer.
Since the Ta $5d$ electrons predominately occupy $d_{3z^2-r^2}$ orbit near the Fermi level,
forming an effective single orbital system. Therefore the $LS$ coupling plays
little role on the groundstate electronic properties on $1$T-TaS$_{2}$ monolayer and bulk,
though its coupling strength is considerable. It is also worthy of stressing that the
the orbital properties of bulk 1T-TaS$_{2}$ are expected to similar to the present orbital
character of monolayer, as seen the {\it Supplementary Materials}.

One may notice that $1$T-TaS$_{2}$ is a strongly coupled e-ph interacting system
\cite{JPCM23.213001}, we anticipate that the electron-phonon interaction may play
considerable roles not only in the star-of-David distortion, but also in transport
in monolayer, as the polaronic effect in dynamic process of bulk\cite{Science344.177}.
The successful synthesis of $1$T-TaS$_{2}$ monolayer may pave a new way to
study the novel properties of 1T-TaS$_{2}$ monolayer with CDW state and Mott physics;
its flat band character and insulating gap could lead to exciton-polaron coupling,
exciton luminescence and dynamical properties, which may provide the possible
applications of transition-metal dichalcogenides in electronic devices and sensors.
%
%

The authors acknowledge X. H. Chen for providing primary experimental data.
This work was supported by the National Science Foundation of China under Grant no.
$ 11274310$ and $11104274$, and the Hefei Center for Physical Science and Technology
under Grant no. 2012FXZY004. Numerical calculations were performed at
the Center for Computational Science of CASHIPS.


\begin{widetext}
\linespread{2}
\begin{center}
\begin{table*}[!hbp]

\caption{Orbital resolved character of a Ta star-of-David  in distorted monolayer $1$T-TaS$_{2}$.}
\begin{tabular}{c|cccccccc}
\hline
\hline
\hline
\hline
    & \ \ \ \ $13$ Ta atoms \ \ \ \ & \ \ \ \  $26$ S atoms  \ \ \ \ & \ \ \ \ Interstitial electrons\\
\hline
Atomic weight & $0.47925$ & $0.18420$ & $0.33654$ \\
\hline
\hline
Central atom & \ \ \ \ $1$ Ta \ \ \ \ & \ \ \ \ $5d$ orbital \ \ \ \ & \ \ \ \ $5d_{3z^{2}-{r}^2}$  orbital\ \ \ \ & \ \ \ \ Other $5d$ orbitals\\
\hline
Orbital weight & $0.14125$ & $0.14090$ & $0.13917$ & $0.00174$ \\
\hline
\hline
First-ring atoms & \ \ \ \ $6$ Ta \ \ \ \ & \ \ \ \  $5d$ orbital \ \ \ \ & \ \ \ \ $5d_{3z^{2}-{r}^2}$  orbital\ \ \ \ & \ \ \ \ Other $5d$ orbitals \\
\hline
Orbital weight & $0.24214$ & $0.03983$ & $0.03667$ & $0.00316$ \\
\hline
\hline
Second-ring atoms & \ \ \ \ $6$ Ta \ \ \ \ & \ \ \ \  $5d$ orbital \ \ \ \ & \ \ \ \ $5d_{3z^{2}-{r}^2}$ orbital \ \ \ \ & \ \ \ \ Other $5d$ orbitals \\
\hline
Orbital weight & $0.09587$ & $0.01581$ & $0.00174$ & $0.01407$ \\
\hline
\hline

\end{tabular}
\end{table*}
\end{center}
\end{widetext}

\end{document}